\documentstyle[12pt]{article}
\newcommand{\alpr}{{\alpha^{\prime}}}
\newcommand{\pr}{\prime}
\newcommand{\pa}{\partial}

\newcommand{\ma}{\mathcal}

\newcommand{\fr}{\frac}

\newcommand{\no}{\nonumber \\}

\newcommand{\ep}{\epsilon}
\newcommand{\beq}{\begin{equation}}
\newcommand{\eeq}{\end{equation}}
\newcommand{\beqa}{\begin{eqnarray}}
\newcommand{\eeqa}{\end{eqnarray}}
\newcommand{\al}{\alpha}

\newcommand{\de}{\delta}

\newcommand{\Ga}{\Gamma}

\newcommand{\ps}{\psi}
\newcommand{\ops}{\overline{\psi}}
\newcommand{\leripa}{{\mathop{\partial}^{\leftrightarrow}}}
\newcommand{\oep}{\overline{\epsilon}}
\newcommand{\ri}{\rightarrow}

\begin{document}
\begin{titlepage}
\thispagestyle{empty}
\begin{flushright}
UT-931 \\
hep-th/0104176 \\
April, 2001 
\end{flushright}

\bigskip
\bigskip

\begin{center}
\noindent{\Large \textbf{On the Supersymmetry of Non-BPS D-brane}}\\
\vspace{2cm}
\noindent{Seiji Terashima\footnote{seiji@hep-th.phys.s.u-tokyo.ac.jp} 
and Tadaoki Uesugi\footnote{uesugi@hep-th.phys.s.u-tokyo.ac.jp} }\\

\bigskip
{\it Department of Physics, Faculty of Science \\ University of Tokyo \\
\medskip
Tokyo 113-0033, Japan}

\end{center}
\begin{abstract}
In this paper we extend the boundary string field theory action for a
 non-BPS D-brane to the one including the target space fermions and the 
nonlinear supersymmetry with 
32 supercharges up to some order. This is based on the idea that the vacuum with a non-BPS 
D-brane belongs to the spontaneously broken phase of the supersymmetry.
As a result, we find that the action is almost uniquely determined up 
to the field redefinition ambiguities.  
\end{abstract}
\end{titlepage}

\section{Introduction}

The supersymmetry is one of the most important subjects to know the full 
structure of string theory. In fact the tremendous progress of understanding 
the non-perturbative aspects of string theory has relied on the supersymmetry. 
In RNS formalism such a ten dimensional supersymmetry in 
various supersymmetric string theories is induced by the superconformal 
invariance and GSO-projection on the string sigma model action. However, we do 
not understand why the supersymmetry naturally appears in the string theory 
and why such prescriptions make it possible to introduce the supersymmetry. 
What is the connection between two dimensional superconformal symmetry and ten 
dimensional supersymmetry? What is the mathematical and the physical meanings 
of GSO-projections?  

On the other hand, recently there has been a lot of studies (for example, see 
\cite{sen18, Gab1}) on the non-BPS physics with different motivations from the 
above arguments, and the off-shell dynamics of non-BPS systems has been 
possible to discuss. The off-shell actions on non-BPS systems are the most 
important quantities which can be obtained from the various string field 
theories. Especially by using boundary string field theory 
\cite{Wi2, Ma, NiPr}, the exact actions for non-BPS D-branes and 
brane-antibrane systems has been recently computed 
\cite{KuMaMo2, Ts2, KrLa, TaTeUe}. Unfortunately 
the string field theories including the R-sector have not been found until 
now, therefore these actions are written only by the bosonic fields. 

{}From these actions we can understand various dynamical aspects of non-BPS 
systems. For example, a non-BPS D-brane decays to the closed string vacuum 
by the trivial tachyon condensation or to a BPS D8-brane by the one with
the topological defect. This fact means that the vacuum 
with the supersymmetry is connected to the vacuum with the non-BPS D-brane.
Then it is natural to consider that the vacuum with non-BPS D-branes 
belongs to the spontaneously broken phase of the supersymmetry in type II 
theories \cite{sen22, Yo}. If so, in the field 
theory on a non-BPS D-brane the nonlinear supersymmetry should be
realized. This is a interesting subject because in non-gravitational field 
theory with 32 supercharges the linear supersymmetry is not known. 
Moreover there is no GSO-projection on the fields which are excited by open 
strings, thus from this we may understand the relation between GSO-projection 
and the supersymmetry.

The supersymmetry with 32 supercharges is very large, thus we expect that only 
this symmetry gives enough constraints to determine the action almost 
completely. Then, as a first step to understand this, we try to extend the 
field theory action of the boundary string field theory on a non-BPS D-brane 
to the one with target space fermions which has the non-linear supersymmetry.

In the next section we review the present understandings of the supersymmetry 
on a non-BPS D-brane, and in section 3 we explicitly construct the field 
theory action with nonlinear supersymmetry by considering all possible terms 
in the action, supersymmetry transformations and field redefinitions up to 
some order. This action is unique up to the field redefinition ambiguities.
The conclusion and the discussion are in section 4.

\section{Supersymmetry on a Non-BPS D-brane}  

First we review the various arguments about the supersymmetry on a non-BPS 
D-brane. In this paper we consider the non-BPS D9-brane in type IIA 
superstring theory. If we need to consider the non-BPS D$p$-brane, we 
have only to perform the T-duality, therefore it is enough to consider the 
D9-brane. 

On a non-BPS D9-brane there are infinite numbers of the fields which are 
excited by the open strings without GSO-projection. They are massive fields, 
a U(1) gauge field, a non-chiral Majorana fermion of SO(1,9) and a real 
tachyon. The tachyon and the Majorana fermion belong to the adjoint 
representation for the U(1) gauge transformation, that is, they are 
gauge-singlets. Formally we can define a effective action written by all these fields on the 
non-BPS D-brane, however in general for our convenience we integrate out the 
massive fields which we do not need to consider\footnote{"Integrate out" means 
by inserting the solution to the equation of motion of the field because here 
we consider the effective action only at the {\sl tree level}.}. 

In \cite{sen22} the author gives the action with the massless fermion in 
which the tachyon is integrated out. Its action is \footnote{In 
this paper we set $2\pi\alpr$ to 1 for simplicity.}
\beqa
\label{Senaction}
S&=&-\sqrt{2}T_9\int d^{10}x \sqrt{-\det(\eta_{\mu\nu}+{\ma F}_{\mu\nu})},\no
&=&-\sqrt{2}T_9\int d^{10}x~[1+\fr{1}{4}F^{\mu\nu}F_{\mu\nu}
-\ops\Ga^{\mu}\pa_{\mu}\ps
+F^{\mu\nu}(\ops\Ga_{11}\Ga_{\mu}\pa_{\nu}\ps)+\cdots],
\eeqa
where
\beqa
{\ma F}_{\mu\nu}&=&F_{\mu\nu}-(\ops\Ga_{\nu}\pa_{\mu}\ps)
-(\ops\Ga_{\mu}\pa_{\nu}\ps)\no
&&-(\ops\Ga_{11}\Ga_{\nu}\pa_{\mu}\ps)
+(\ops\Ga_{11}\Ga_{\mu}\pa_{\nu}\ps)
+(\ops\Ga^{\rho}\pa_{\mu}\ps)(\ops\Ga_{\rho}\pa_{\nu}\ps)\no
&&+\fr{1}{2}
(\ops\Ga_{11}\Ga_{\rho}\pa_{\mu}\ps)(\ops\Ga^{\rho}\pa_{\nu}\ps)
-\fr{1}{2}(\ops\Ga_{11}\Ga_{\rho}\pa_{\nu}\ps)(\ops\Ga^{\rho}\pa_{\mu}\ps).
\eeqa
Here we write the effective action whose diffeomorphism invariance is already 
fixed by the static gauge \cite{sen22}. 

However, the most characteristic property of this action is that this action 
has the {\sl nonlinear} supersymmetry with 32 supercharges. The 
transformations are as follows:
\beqa
\label{Sensusytr}
\de \ps_{\al}&=&\ep_{\al}-(\oep\Ga^{\mu}\ps)(\pa_{\mu}\ps)_{\al},\no
\de A_{\mu}&=&(\oep\Ga_{11}\Ga_{\mu}\ps)
-(\oep\Ga^{\nu}\ps)\pa_{\nu}A_{\mu}
-(\oep\Ga^{\rho}\pa_{\mu}\ps)A_{\rho}\no
&&-\fr{1}{6}\{(\oep\Ga_{11}\Ga_{\nu}\ps)(\ops\Ga^{\nu}\pa_{\mu}\ps)
+(\oep\Ga_{\nu}\ps)(\ops\Ga_{11}\Ga^{\nu}\pa_{\mu}\ps)\}.
\eeqa 
{}From these transformations we can obtain the following supersymmetry algebra:
\beqa
\label{Sensusyps}
[\de_1,~\de_2]\ps_{\al}=2(\oep_1\Ga^{\rho}\ep_2)\pa_{\rho}\ps_{\al}~~~,~~~
[\de_1,~\de_2]A_{\mu}=-2(\oep_1\Ga_{11}\Ga_{\mu}\ep_2)
+2(\oep_1\Ga^{\rho}\ep_2)\pa_{\rho}A_{\mu}.
\eeqa  
Thus, we can obtain the expected translations together with a constant shift 
of the gauge field which can be regarded as an irrelevant gauge 
transformation. This nonlinear supersymmetry realizes the idea 
that the vacuum with a non-BPS D-brane belongs to the spontaneously broken 
phase of the 32 supersymmetries of type II theories. The evidence of the 
action (\ref{Senaction}) and the supersymmetry transformation 
(\ref{Sensusytr}) is given in \cite{HaYo} with the explicit calculations of 
the on-shell scattering amplitudes. 

The configurations where this supersymmetry restores are the closed string 
vacuum or the vacuum with BPS D-branes. At these configurations the total or 
partial supersymmetries are linearly represented. These stable vacuum and 
unstable one with the non-BPS D-brane are connected by the process of the tachyon condensation. Therefore if we want to consider the off-shell dynamics of 
non-BPS D-branes and realize this supersymmetry at the arbitrary off-shell 
point, we have to incorporate the tachyon field. Namely, we need the effective 
action before the tachyon is integrated out. Therefore from (\ref{Senaction}) 
we can not see the mechanism of the restoration of the supersymmetry.

On the other hand, recently the effective action which is represented only by 
the bosonic fields including the tachyon ($\ps(x)$ is set to zero) has been 
obtained by the open string field theory. Especially, some exact results on 
the effective action of the various non-BPS systems have been derived from 
the boundary string field theory \cite{KuMaMo2, Ts2, KrLa, TaTeUe}. 
The action for a non-BPS D-brane \cite{An, KuMaMo2} is  
\beqa
\label{BSFTaction}
S&=&-\sqrt{2}T_9\int d^{10}x e^{-\fr{1}{4}T^2}
\frac{\prod^{\infty}_{r=\fr{1}{2}}\det(\eta_{\mu\nu}+F_{\mu\nu}
+\fr{1}{4\pi r}\pa_{\mu}T\pa_{\nu}T)}{\prod^{\infty}_{m=1}\det(m\eta_{\mu\nu}
+mF_{\mu\nu}+\fr{1}{4\pi}\pa_{\mu}T\pa_{\nu}T)},\no
&&\no
&=&-\sqrt{2}T_9\int d^{10}x e^{-\fr{1}{4}T^2}\sqrt{-\det(\eta_{\mu\nu}
+F_{\mu\nu})}{\ma F}\left[\fr{1}{4\pi}{\ma G}^{\mu\nu}\pa_{\mu}T\pa_{\nu}
T\right],
\eeqa
where
\beqa
{\ma G}^{\mu\nu}\equiv\left(\fr{1}{1+F}\right)^{(\mu\nu)}~~~,~~~
{\ma F}[x]\equiv\fr{4^x x (\Ga(x))^2}{2\Ga(2x)}=1+2(\log 2)x+O(x^2).
\eeqa  
Here, $(\mu\nu)$ indicates the symmetrization. 
The ``exactness'' of this action means that it is obtained by integrating out 
{\sl all} massive fields\footnote{In the other string field theory it is 
difficult to perform this, therefore the approximation is used which is 
usually called level truncation.}with all $\alpr$ corrections under the 
conditions that $\pa_{\mu}\pa_{\nu}T=0,$ and $\pa_{\rho}F_{\mu\nu}=0$. 
\footnote{However, even if we regard (\ref{BSFTaction}) as the complete action 
without this condition, the action seems to give several nontrivial results. 
For example, in \cite{HaHi} the author constructed the various soliton 
solutions which do {\sl not} obey the above conditions.} 

As an example of the tachyon condensation in (\ref{BSFTaction}) we consider 
the following nontrivial exact soliton solutions which represents a BPS 
D8-brane localized at $x^9=0$ after the tachyon condensation \cite{KuMaMo2}:
\beqa
\label{solution}
T(x)=ux^9~~~ (u\rightarrow\infty)~~~,~~~A_{\mu}(x)=0~~~,~~~\ps(x)=0.
\eeqa     
This solution gives a correct tension and a D8-brane charge from the action 
(\ref{BSFTaction}) and Wess-Zumino term of the non-BPS D-brane 
\cite{KrLa, TaTeUe}. These facts give the evidences for the identification of 
this solution as a BPS D8-brane.

{}From these two actions (\ref{Senaction}) and (\ref{BSFTaction}) we expect 
that in boundary string field theory action (\ref{BSFTaction}) the nonlinear 
supersymmetry can be realized by including the transformations not only of the 
gauge field and the fermion but also of the tachyon. Indeed we can 
prove the existence of such an action and supersymmetry transformations. First 
we define $S_0(C^a,~\lambda^i)$ as an action where not only  
$T(x),~\ps(x)~\mbox{and}~A_{\mu}(x)$ (here, we represent these three fields as 
$C^a(x)~(a=1,2,3))$ but also the massive fields $\lambda^i(x)$ are included. 
The supersymmetric invariance of the action is written by
\beqa
\label{S0}
S_0(C^a+\de C^a,~\lambda^i+\de\lambda^i)=S_0(C^a,~\lambda^i)~~~
\mbox{where}~~~\left\{\begin{array}{rcl}
\de C^a&=&\de C^a (C^a,\lambda^i),\\
\de \lambda^i&=&\de \lambda^i(C^a,\lambda^i).
\end{array}\right.
\eeqa
Then, we integrate out the massive fields $\lambda^i(x)$ from 
$S_0(C^a,~\lambda^i)$. This is performed by inserting the solution to the 
equation of motion for $\lambda^i(x)$. If we define $\bar{\lambda}^i=
\bar{\lambda}^i(C^a)$ as the solution to the equation of motion 
for $\lambda^i(x)$ and insert this solution into (\ref{S0}), then the result is
\beqa
S_0(C^a,~\bar{\lambda}^i)&=&S_0(C^a+\de C^a(C^a,\bar{\lambda}^i),
~\bar{\lambda}^i+\de \lambda^i(C^a,\bar{\lambda}^i))\no
&=&S_0(C^a+\de C^a(C^a,\bar{\lambda}^i),~\bar{\lambda}^i).
\eeqa
Here we used the relation 
$\fr{\de S_0}{\de \lambda^i}|_{\lambda=\bar{\lambda}}=0$. If we define the 
action in which the massive fields are integrated out as $S(C^a)\equiv 
S_0(C^a,\bar{\lambda}^i(C^a))$, the above relation indicates that $S(C^a)$ is 
invariant for $C^a\rightarrow C^a+\bar{\de} C^a$ where 
$\bar{\de} C^a(C^a)\equiv \de C^a(C^a,\bar{\lambda}^i(C^a))$. 

This conjecture of the realization of the nonlinear supersymmetry with 32 
supercharges is highly nontrivial because, regardless of the dimension, 
in the supersymmetric field theory with 32 supercharges any supermultiplets 
can not be constructed without introducing the graviton. This indicates that 
we can not construct any field theories with the {\sl linear} supersymmetry. 
However, the {\sl nonlinear} realization of the supersymmetry with 32 
supercharges is an exception we can consider to avoid this no-go 
theorem. 

In (\ref{Sensusytr}) the supersymmetry 
transformations have been already obtained,
however, from these equations we 
can not discuss the tachyon condensation. If we can obtain the action 
without integrating out the tachyon field, we can understand the mechanism of 
the restoration of the linear supersymmetry through the analysis of the 
tachyon condensation. 
 
The idea of the realization of the nonlinear supersymmetry in the off-shell 
region was considered in \cite{Yo} by using Witten's 
cubic type open superstring field theory \cite{Wisu}. In this paper the author 
generalized this theory to the string field theory on the non-BPS D-brane by 
including the GSO-odd components of the string field, and constructed the 
nonlinear supersymmetry transformations of the string field. However, Witten's 
superstring field theory has a problem about the collision of picture changing 
operators, thus this discussion is rather formal. Moreover in this theory 
the explicit BPS solution has been not obtained until now, therefore we can 
not see how the linear supersymmetry restores at this stable vacuum. 

On the other hand, contrary to this theory, the recent work on the
boundary string field theory has given the exact action
(\ref{BSFTaction}) and the exact soliton solution (\ref{solution}). 
Therefore by using these results we expect that the nonlinear 
supersymmetry can be realized explicitly in the boundary string field
theory action (\ref{BSFTaction}) which will become the action 
(\ref{Senaction}) with 
integrating out the tachyon. If so, we can see the restoration of the linear 
supersymmetry after the tachyon condensation. Concretely speaking, the 
solution (\ref{solution}) is
expected to be a BPS solution in the sense of the supersymmetric field
theory, that is, this solution will satisfy the BPS equation $\de_{SUSY} 
\ps(x)=0$. If we expand the action around this solution and
integrate out the massive fields, the result is expected to become the 
supersymmetric Dirac-Born-Infeld action for a BPS D-brane
\cite{AgPoSc, ArFrThTs}. 

Of course, without the knowledge of the string field theory with R-sector, it 
may be difficult to obtain the full action which includes all terms with 
fermions. However, in the supersymmetric quantum mechanics (for example, see 
\cite{CoKhSu}) there is an interesting example of the model:
\beqa
\label{onedim}
S=\int dt~e^{-\fr{T^2}{4}}\left[\fr{1}{2}(\pa_t T)^2-1
+\fr{i}{2}\ps^+\leripa_t\ps^-+\fr{T}{2\sqrt{2}}\ps^+ \ps^- \right].
\eeqa
And this action is invariant by the following {\sl nonlinear} supersymmetry 
transformations:
\beqa
\!\!\!   \de T\!\!  &=& \!i (\ps^+\eta^-+\ps^-\eta^+),\no
\!\!\! \de\ps^+ \!\! &=& \!\!\! -i\sqrt{2}\eta^+ -\pa_t T\eta^+
+\fr{i}{4}T\ps^+\ps^-\eta^+\!\!\!,
\;\;
\de\ps^-=i\sqrt{2}\eta^- -\pa_t T\eta^- \! -\fr{i}{4}T\ps^+\ps^-\eta^-,
\eeqa
and these transformations satisfy the usual supersymmetry algebra.
This is equivalent to a well known action in supersymmetric 
quantum mechanics:
\beqa
S=2\pi\int dt \left[\fr{1}{2}\dot{\phi}^2-\fr{1}{2}W^2(\phi)
-W^{\prime}(\phi)\theta^{+}\theta^{-}+i\theta^{+}\dot{\theta}^{-}\right].
\eeqa 
\beqa
\label{onedimtr}
\de \phi&=&i (\theta^{+}\eta^{-}+\theta^{-}\eta^{+}),
\;\;\;\;\;\;\;
\de \theta^{\pm}=\mp i(W(\phi)\mp i\dot{\phi}) \eta^{\pm},
\eeqa
where $\phi=\fr{2}{\sqrt{\pi}}\int^{\fr{T}{2\sqrt{2}}}_0 e^{-s^2}ds,
~W(\phi)=\fr{1}{\sqrt{\pi}}\exp{(-\fr{1}{8}T(\phi)^2)}$ and 
$\theta^{\pm}=(2\pi)^{-\fr{1}{2}}e^{-\fr{T^2}{8}}\ps^{\pm}$. 
This action shows that the nonlinear supersymmetry can be included for the 
action with the boundary string field theory potential $\exp(-\fr{T^2}{4})$ at 
least in one dimension. At $T=\pm\infty$ the superpotential vanishes, that
is $W=0$, and the supersymmetry restores as we can see from
(\ref{onedimtr}). 
Moreover this has BPS solutions $T=\pm\sqrt{2}t_E$ if we perform 
the Wick rotation $t~\rightarrow-it_E$\footnote{It is not essential to 
consider the Euclidean action. Indeed, we will construct an analog of this 
model in four dimension in appendix A.}. 

Of course the extension of this model to the one in ten dimension is 
nontrivial because the field theory on a non-BPS D-brane has infinite number 
of terms and because in (\ref{onedim}) the gauge field on a non-BPS D9-brane 
is not included. However, there is a truncated model which approximately 
describes the tachyon condensation on a non-BPS D-brane \cite{MiZw2}. 
The action is
\beqa
S=-\sqrt{2}T_9\int d^{10}x~e^{-\fr{1}{4}T^2}
\left[\fr{1}{2}\pa^{\mu}T\pa_{\mu}T+1+\fr{1}{4}F^{\mu\nu}F_{\mu\nu}
-\ops\Ga^{\mu}\pa_{\mu}\ps+\fr{1}{2\sqrt{2}}T\ops\ps\right].
\eeqa
The action (\ref{onedim}) is obtained by the dimensional reduction and
truncations of some fields from this one. 
In \cite{MiZw2} by expanding the action around the kink solution
\beqa
\label{solution2}
T(x)=\sqrt{2}x^9~~~,~~~A_{\mu}(x)=0~~~,~~~\ps(x)=0,
\eeqa 
which is similar to (\ref{solution}), the authors investigate the spectrum on
this soliton. Note that the $T\ops\ps$ term is like a mass term of domain
wall type to produce chiral fermion. As a result the massless
spectrum is 
the vector supermultiplet in 9 dimension, which is same as the one on the BPS 
D8-brane. The massive spectrums are a little different, while there are no 
continuous spectrums and the mass square differences are equal spacing, 
which is the similar structure as the one in string theory. However, in 
\cite{MiZw2} the supersymmetry is not considered, thus we can not understand 
how the linear supersymmetry restores at a stable vacuum.     

In this paper as a first step to understand the full structure of the 
nonlinear supersymmetry on a non-BPS D-brane, we will consider the
several leading terms of the effective action and will confirm that the nonlinear supersymmetry 
is realized and that the action is consistent with the result from
boundary string field theory. 
 
\section{Construction of the Action with Nonlinear Supersymmetry}

In this section we will analyze the constraints which come from the 
supersymmetric invariance of the action and will show that the nonlinear 
supersymmetry can be realized on the effective action 
of a non-BPS D9-brane. 

First we consider the most general action and the supersymmetry 
transformation which are constructed by $T(x), A_{\mu}(x)$ and $\ps(x)$ 
up to several order. Here we pick up the terms which satisfy the following 
conditions:
\beqa
\#A_{\mu}(x)=0,~~ \#\ps(x)=1~~ \mbox{and}~~ \#\pa_{\mu}\leq 1~~~
\mbox{in}~\de S,  
\eeqa
where $ \# X$ is the number of $X$ and $\de S$ is the variation of the
action by 
the supersymmetry transformation. Of course, the following analysis can be 
done by using the $\alpr$ expansion in the same way.
Under this condition the relevant terms are as follows
\beqa
\label{action1}
S&=&-\sqrt{2}T_9\int d^{10}xe^{-\fr{1}{4}T^2}\no
&&\times [1+f_1(T)(\ops\ps)+ f_2(T)(\ops\Ga^{\mu}\pa_{\mu}\ps)
+f_3(T)(\ops\Ga_{11}\Ga^{\mu}\pa_{\mu}\ps)].
\eeqa
And we also consider the supersymmetry transformations
\beqa
\label{susyps1}
\de\ps_{\al}&=&g_1(T)\ep_{\al}+g_2(T)(\Ga_{11}\ep)_{\al}
+g_3(T)\pa^{\mu}T(\Ga_{\mu}\ep)_{\al}+g_4(T)\pa^{\mu}T
(\Ga_{11}\Ga_{\mu}\ep)_{\al},\\
\label{susyt1}
\de T&=&h_1(T)(\oep\ps)+h_2(T)(\oep\Ga_{11}\ps)
+h_3(T)\pa^{\mu}T(\oep\Ga_{\mu}\ps)\no
&&+h_4(T)\pa^{\mu}T
(\oep\Ga_{11}\Ga_{\mu}\ps)
+h_5(T)(\oep\Ga^{\mu}\pa_{\mu}\ps)
+h_6(T)(\oep\Ga_{11}\Ga^{\mu}\pa_{\mu}\ps),
\eeqa
where the coefficient functions $f_i(T),~g_i(T)$ and $h_i(T)$ are nonsingular 
around $T=0$. Now we will consider the constraints which come from the 
supersymmetric invariance of the action (\ref{action1}), that is, 
$\delta S=0$. These are the relations between the coefficient functions: 
\beqa
\label{delS1}
\de S&=&-\sqrt{2}T_9\int d^{10}x e^{-\fr{1}{4}T^2}\no
&&\times[C_1(T)(\oep\ps)+C_2(T)(\oep\Ga_{11}\ps)
+C_3(T)\pa^{\mu}T(\oep\Ga_{\mu}\ps)
+C_4(T)\pa^{\mu}T(\oep\Ga_{11}\Ga_{\mu}\ps)],\no
\eeqa
where
\beqa
&&\begin{array}{rcl}
C_1&=&-\fr{1}{2}T h_1+2f_1g_1~~~,~~~C_2=-\fr{1}{2}T h_2-2f_1g_2,\no
&&\no
C_3&=&\fr{1}{2}Tf_2g_1-2f_1g_3-\fr{1}{2}T h_3-2f_2\dot{g}_1
+2f_3\dot{g}_2\no
&&+(\fr{1}{2}-\fr{1}{4}T^2)h_5-\fr{1}{2}Tf_3g_2+\fr{1}{2}T\dot{h}_5
-\dot{f}_2g_1+\dot{f}_3g_2,\no
&&\no
C_4&=&-\fr{1}{2}Tf_2g_2-2f_1g_4-\fr{1}{2}T h_4+2f_2\dot{g}_2
-2f_3\dot{g}_1\no
&&+(\fr{1}{2}-\fr{1}{4}T^2)h_6+\fr{1}{2}Tf_3g_1
+\fr{1}{2}T\dot{h}_6+\dot{f}_2g_2-\dot{f}_3g_1,
\end{array}
\eeqa
and $\dot{f}$ denotes $df/dT$. {}From this we can obtain four constraint 
equations $C_i(T)=0~(i=1,2,3,4)$.

However, only with these conditions we can not completely fix the functions 
$f_i(T), g_i(T)$ and $h_i(T)$. In general the effective action and the 
transformations of fields which represent some symmetries have the 
field redefinition ambiguities of the realization. Of course the field 
redefinition does not change the physics. Especially in the boundary
string field theory this ambiguity comes from the regularization of the 
nonlinear sigma model. Therefore also in this case we have to consider
the most general field redefinition of $T(x)$ and $\ps(x)$ which are 
nonsingular around $T=0$, and fix these ambiguities.

The field redefinitions which are needed in (\ref{action1}), (\ref{susyps1}) 
and (\ref{susyt1}) are 
\beqa
\label{fieldps1}
\ps^{\pr}&=&k_1(T)\ps_{\al}+k_2(T)(\Ga_{11}\ps)_{\al}
+k_3(T)\pa^{\mu}T(\Ga_{\mu}\ps)_{\al}
+k_4(T)\pa^{\mu}T(\Ga_{11}\Ga_{\mu}\ps)_{\al}\no
&&+k_5(T)(\Ga^{\mu}\pa_{\mu}\ps)_{\al}
+k_6(T)(\Ga_{11}\Ga^{\mu}\pa_{\mu}\ps)_{\al},\\
\no
\label{fieldt1}
T^{\pr}&=&T+l_1(T)(\ops\ps)+l_2(T)(\ops\Ga^{\mu}\pa_{\mu}\ps)
+l_3(T)(\ops\Ga_{11}\Ga^{\mu}\pa_{\mu}\ps).
\eeqa
Moreover the parameter $\ep_{\al}$ of the supersymmetry transformation in 
(\ref{susyps1}) and (\ref{susyt1}) has the following ambiguity:
\beqa
\ep_{\al}~\rightarrow~ a\ep_{\al}+b(\Ga_{11}\ep)_{\al},
\eeqa
where $a,b$ are constants. We have fixed the first term of eq(\ref{fieldt1}) 
to $T$, not general function of $T$. This is because $T\rightarrow f(T)$ 
changes the potential. We have fixed the potential to $e^{-\fr{1}{4}T^2}$ which was 
obtained by boundary string field theory \cite{KuMaMo2, Ts2}. 

{}From (\ref{action1}), (\ref{susyps1}) and (\ref{susyt1}) we can see that 
there are 13 coefficient functions $f_i(T),~g_i(T)$ and $h_i(T)$ which are 
not determined. On the other hand (\ref{delS1}) gives 4 constraint equations 
which come from the supersymmetry, and (\ref{fieldps1}) and (\ref{fieldt1})
represent the 9 freedoms of the field redefinitions. Therefore we expect 
that all coefficient functions $f_i(T), g_i(T)$ and $h_i(T)$ can be determined 
completely. 

In fact we can do. At first order the coefficients $g_1(T)$ and $g_2(T)$ in 
eq(\ref{susyps1}) transforms as follows:
\beqa
\label{amb1}
g_1~\ri~k_1g_1+k_2g_2~~~,~~~
g_2~\ri~k_1g_2+k_2g_1.
\eeqa
{}From these we can consistently fix $g_1(T)$ and $g_2(T)$ to 1 and 0, 
respectively. By representing the action and the supersymmetry transformations 
with the redefined fields $\ps^{\pr}(x)$ and $T^{\pr}(x)$, the coefficient 
functions $f_i(T),~ g_i(T)$ and $h_i(T)$ transform as follows:\footnote{
Sting theory should keep G-parity invariance before and after 
interactions. From this we can restrict all the coefficient functions 
$f_i(T), g_i(T),h_i(T), k_i(T)$ and $l_i(T)$ to the even or odd functions
 of T.}
\beqa
\label{amb2}
f_1~&\ri&~f_1+\fr{1}{2}T l_1,\no
f_2~&\ri&~f_2-2f_1k_5+\fr{1}{2}T l_2-Tl_1k_5~~~,~~~
f_3~\ri~f_3-2f_1k_6+\fr{1}{2}T l_3-Tl_1k_6,\no
g_3~&\ri&~g_3+k_3~~~,~~~g_4~\ri~g_4+k_4~~~,~~~h_1~\ri~h_1+2l_1~~~,~~~
h_2~\ri~h_2,\no
h_3~&\ri&~h_3-2l_1g_3-k_3h_1-2k_3l_1-k_4h_2,\no
h_4~&\ri&~h_4-2l_1g_4-k_4h_1-2k_4l_1-k_3h_2,\no
h_5~&\ri&~h_5-k_5h_1-2k_5l_1-k_6h_2~~~,~~~h_6~\ri~h_6-k_6h_1-2k_6l_1-k_5h_2.
\eeqa
Therefore from these equations we can consistently 
fix the coefficient functions and obtain the following results:
\footnote{Strictly speaking if we fix $h_1(T)$ to zero, we can not fix 
$h_3(T), h_4(T), h_5(T)$ and $h_6(T)$. 
} 
\beqa
\label{action4}
S&=&-\sqrt{2}T_9\int d^{10}xe^{-\fr{1}{4}T^2}
[1+\fr{1}{2\sqrt{2}}T(\ops\ps)-(\ops\Ga^{\mu}\pa_{\mu}\ps)],\no
&&\no\label{susyMi0}
\de\ps_{\al}&=&\ep_{\al}+\fr{1}{\sqrt{2}}\pa^{\mu}T(\Ga_{\mu}\ep)_{\al}~~,
~~\de T~=~\sqrt{2}(\oep\ps)-2\pa^{\mu}T(\oep\Ga_{\mu}\ps).
\eeqa

{}From this result we find that we can obtain the same action 
as the one considered in \cite{MiZw2} by appropriate fixing of 
the field redefinition ambiguities. Also from $\de\ps(x)=0$ we can obtain the 
same 1/2-BPS solutions as (\ref{solution2}). For this solution the 
1/2 supersymmetry which satisfy $\Ga^9\ep=-\ep$ restores on the BPS D-brane. 
If we again redefine the new field in order to normalize the kinetic
term to 1, then the exponential potential appears in front of the
right-hand side of (\ref{susyMi0}) and we obtain another BPS solution
$T\rightarrow\pm\infty$ which represents the closed string vacuum in the 
same way as (\ref{onedimtr}). The above result indicates that the structure 
of the supersymmetry can be included in the action of \cite{MiZw2}. 

{}From this explicit practice we obtained the effective action with the 
nonlinear supersymmetry at the leading order. However in this order the gauge 
fields are not introduced. Therefore we will extend this result to 
the effective action with more terms which satisfy the following conditions. 
\beqa
\label{condition2}
\#A_{\mu}(x)\leq1,~~ \#\ps(x)=1~~ \mbox{and}~~ \#\pa_{\mu} \leq 2~~
\mbox{in}~\de S, 
\eeqa
where $ \# X$ is the number of $X$.
In the appendix C we write the 
general action, the supersymmetry transformations and the field 
redefinitions up to the above order. The process to determine the coefficient 
functions of the action and the supersymmetry transformations is same as the 
previous one, therefore we abbreviate this process. We fix the field 
redefinition ambiguities appropriately. The final form of the action is
\footnote{The fact is that the total number of the constraints from the 
supersymmetric invariance and the field redefinition ambiguities 
(\ref{fieldps3}) and (\ref{fieldt3}) is a little more than the freedoms which 
appear in the action (\ref{action3}) and the supersymmetry transformations 
(\ref{susyps3}), (\ref{susyt3}) and (\ref{susyA3}). Nevertheless we find
that various coefficient functions can be fixed with keeping the 
supersymmetry.}\footnote{We can change the coefficient of the term
$\pa^{\mu}T\pa_{\mu}T$ in the action by fixing the field 
redefinition ambiguities differently. Therefore it is also possible to
take another fixing to produce the boundary string field theory action 
(\ref{BSFTaction}).}
\beqa
\label{higheraction}
S&=&-\sqrt{2}T_9\int d^{10}xe^{-\fr{1}{4}T^2}\no
&&\times [1+\fr{1}{2}(\pa^{\mu}T\pa_{\mu}T)+\fr{1}{2\sqrt{2}}T(\ops\ps)
+\fr{1}{4}F^{\mu\nu}F_{\mu\nu}-(\ops\Ga^{\mu}\pa_{\mu}\ps)\no
&&-2\sqrt{2}\pa^{\mu}T(\ops\pa_{\mu}\ps)
+F^{\mu\nu}(\ops\Ga_{11}\Ga_{\mu}\pa_{\nu}\ps)],
\eeqa
and the supersymmetry transformations are
\beqa
\label{susy2}
\de \ps&=&\ep+\fr{1}{\sqrt{2}}\pa^{\mu}T(\Ga_{\mu}\ep)_{\al},\no
\de T&=&\sqrt{2}(\oep\ps)-2\pa^{\mu}T(\oep\Ga_{\mu}\ps)
-\sqrt{2}~\pa^{\mu}T\pa_{\mu}T(\oep\ps),\no
\de A_{\mu}&=&(\oep\Ga_{11}\Ga_{\mu}\ps)
\eeqa
Thus we find that there exists the nonlinear supersymmetry up to this order.
   
Then we have to verify whether these transformations of $\ps(x), 
T(x)$ and $A_{\mu}(x)$ satisfy the supersymmetry algebra. The expected 
forms of the supersymmetry algebra are
\beqa
[ \de_1 , \de_2 ] \ps &=& 2(\oep_1\Ga^{\mu}\ep_2)\pa_{\mu}\ps, \no
{[} \de_1, \de_2 {]} T &=& 2(\oep_1\Ga^{\mu}\ep_2)\pa_{\mu}T, \no
{[} \de_1, \de_2 {]} A_{\mu} &=& -2(\oep_1\Ga_{11}\Ga_{\mu}\ep_2)
+2(\oep_1\Ga^{\rho}\ep_2)\pa_{\rho}A_{\mu}.
\label{susyalg}
\eeqa
We consider that for $\ps(x)$ and $A_{\mu}(x)$ the algebras will be 
same as (\ref{Sensusyps}). Indeed if we repeat supersymmetry 
transformations (\ref{susy2}) again, we can obtain the same result up to the 
reliable order under the condition (\ref{condition2}). For example, the 
translation term $2(\oep_1\Ga^{\mu}\ep_2)\pa_{\mu}\ps$ in $[\de_1,~\de_2]\ps$ 
does not appear from (\ref{susy2}). However, by including the more higher 
terms about fermions in the action (\ref{higheraction}) and the supersymmetry 
transformations (\ref{susy2}) we expect that this term appears correctly. The 
point of this discussion is that the form of the supersymmetry transformation 
is changed by the field redefinitions, while the supersymmetry algebra is 
invariant.\footnote{
There are terms  from the gauge transformation and 
terms proportional to the equations of motion in (\ref{susyalg}).
} On the other hand, if we expand $T(x)$ around the solution 
(\ref{solution}) or (\ref{solution2}) such as
\beqa
\label{expanding}
T(x)=u(x^9-X^9(x^i))
\;\; (u=\infty~ \mbox{or}~ \sqrt{2}~;~i=0,\cdots, 8),
\eeqa 
then the algebra of the $X^9$ becomes the following one:
\beqa
[\de_1,~\de_2]X^9~&=&~-2(\oep_1\Ga^9\ep_2)+2(\oep_1\Ga^i\ep_2)\pa_i X^9.
\eeqa
This algebra coincides with that of the transverse scalar on a BPS
D8-brane \cite{AgPoSc, sen22}. This is a desired result because $X^9(x^i)$ is 
the translation mode of the soliton (BPS D8 brane) \cite{MiZw1,MiZw2} as 
we can see from (\ref{expanding}). 
 
\section{Conclusions and Discussions}

In this paper we introduced the boundary string field theory action and 
constructed the action order-by-order which includes fermions and 
realizes the nonlinear supersymmetry. In the first practice we obtained 
the same action as the Minahan-Zwiebach model \cite{MiZw2} and verified that 
the nonlinear supersymmetry can be realized. Next in order to include
the gauge field, we extend this action to the one with more higher
dimensional  terms. Then we found that we can construct the action with
the nonlinear supersymmetry which is consistent with the boundary string 
field  theory action, and confirmed that this action 
can have the BPS soliton solution in the supersymmetric sense, which is  
similar to the one (\ref{solution}) in the boundary string field theory.

However, any forms of the action are allowed in the above order, therefore 
the definite conclusions are not obtained as far as we consider the finite 
numbers of terms in the action and supersymmetry transformations. 
As an example 
let us consider the BPS solution. {}From the existence of the solution 
(\ref{solution}) in boundary string field theory, we expect that the form of 
$\de \ps(x)$ is the following:
\beqa
\label{generalBPS}
\de \ps_{\al}&=&L(T^2)[M(\pa^{\mu}T\pa_{\mu}T)\ep_{\al}
+N(\pa^{\mu}T\pa_{\mu}T)\pa^{\nu}T(\Ga_{\nu}\ep)_{\al}]\no
&&+(\mbox{terms including}~~\ps,~A_{\mu},~\pa_{\mu}\pa_{\nu}T),
\eeqa
where $L(x),~M(x)$ and $N(x)$ are nonsingular functions around $x=0$, with a relation  
\beqa
\label{limit}
\lim_{x\rightarrow\infty} |xN(x^2)/M(x^2)|=1.   
\eeqa
Note that here the fields are partially fixed according to the boundary string 
field theory, and in that sense (\ref{generalBPS}) depends on the remaining 
field redefinitions. Then if we truncate the above transformation such as 
(\ref{susyps1}), we are able to set the coefficients 
$g_i(T)(i=2,3,4)$ to zero by the field redefinitions (see (\ref{amb1}), 
(\ref{amb2})). {}From this form we can not see the BPS solution explicitly. 
The analysis of truncated actions (\ref{action4}) and 
(\ref{higheraction}) showed that at least the realization of the nonlinear 
supersymmetry does not contradict the existence of the BPS solutions. 

One of the motivations of this paper was from the fact that the truncated 
action in \cite{MiZw2} is a good approximation. However, because the mass of 
the tachyon is the order of the Planck scale, it is difficult to consider the 
reason why such a truncated action can work for the tachyon condensation. Any 
way, to find the full order action with the nonlinear supersymmetry we 
consider that some other algebraic principle to construct the action
should be needed.

As a further study of the supersymmetry, it is interesting to extend the 
analysis in this paper to the non-Abelian non-BPS D-branes because also in 
this case the nontrivial and exact BPS solution is already known
\cite{Te}. There are a lot of things to investigate.

\vskip6mm\noindent
{\bf Acknowledgements}

\vskip2mm
We would like to thank 
T. Hara, K. Ohmori and T. Takayanagi 
for useful discussions.
The works of S.T. 
were supported in part by JSPS Research Fellowships for Young 
Scientists. \\

\appendix
\setcounter{equation}{0}

\section{A Four Dimensional Model with Nonlinear Supersymmetry} 

Here we introduce a four dimensional toy model with the nonlinear supersymmetry
using the four dimensional $N=1$ superfield formulation.
We will use the notation used in \cite{WeBa}.
We consider a chiral superfield 
$\Phi(x,\theta)
= T(x)+ i a(x)+\sqrt{2} \theta \chi +\cdots$ with a Lagrangian
\beq
L=\int d^2 \theta d^2 \bar{\theta } K(\Phi, \bar{\Phi})
+\left[ \int d^2 \theta P(\Phi) + \mbox{h.c.}  \right],
\eeq
where the K\"ahler potential $K$ and the superpotential $P$ are given by
\beq
K(\Phi, \bar{\Phi})= 
\left| \int^\Phi_0 dy e^{-\frac{1}{8} y^2} \right|^2, \,\,\,
P(\Phi)= i \int^\Phi_0 dy e^{-\frac{1}{4} y^2}.
\eeq
Then the metric on the K\"ahler manifold becomes 
\beq
g_{\Phi \bar{\Phi}} = e^{-\frac{1}{8} (\Phi^2+ \bar{\Phi}^2)},
\eeq
$\Gamma^{\Phi}_{\Phi \Phi}= -\frac{1}{4} \Phi$ and 
$R_{\Phi \bar{\Phi} \Phi \bar{\Phi}}=0$.
Then the component Lagrangian is computed as
\beqa
L &=& -e^{-\frac{T^2}{4}} e^{\frac{a^2}{4}} 
\left\{  1+\pa_m T \pa^m T + \pa_m a \pa^m a
+i \bar{\chi} \bar{\sigma}^m \pa_m \chi 
-i \bar{\chi} \bar{\sigma}^m \chi 
\frac{1}{4} (T+ i a) \pa_m (T+ i a)  \right. \nonumber \\
& & \left. \hspace{3cm} 
+i \frac{1}{8} (T+ia) e^{-i \frac{1}{2} a T} \chi \chi 
-i \frac{1}{8} (T-ia) e^{+i \frac{1}{2} a T} \bar{\chi} \bar{\chi}
\right\},
\eeqa
which has a nonlinearly realized supersymmetry at $T=a=0$ given by
\beq
\delta (T+ia)= \sqrt{2} \xi \chi, \;\; 
\delta \chi= i \sqrt{2} \sigma^m \bar{\xi} \pa_m (T+ia)
+\frac{1}{4} (T+i a) \delta(T+ia) \chi 
-i \sqrt{2} e^{i \frac{1}{2} a T} \xi.
\eeq
This model has a 1/2-BPS domain wall solution 
$T= x^3$ and $a=\xi=0$.

If we identify the real scalar $T$ as the tachyon, this action can be regarded 
as a toy model for the tachyon in a non-BPS D-brane.
Indeed, the model has the characteristic properties such as the restoration 
of the supersymmetry at $T=\pm \infty$ where the action vanishes and the 
existence of a 1/2-BPS solution corresponding to a one lower dimensional BPS 
D-brane.

This model can be extended to other dimensions and to 
other matter contents.

\section{SO(1,9) Clifford Algebra} 

The gamma matrices $\Gamma^{\mu}(\mu=0,1,2,\cdots,9)$ are defined by
\beqa
\{\Gamma^{\mu},~\Gamma^{\nu}\}=2\eta^{\mu\nu}.
\eeqa
Here the sign convention of the flat metric $\eta^{\mu\nu}$ is most plus one, 
that is, ${\rm diag}(\eta^{\mu\nu})=(-,+,+ \cdots, +)$.
We define $\Ga_{11}$ as
\beqa
\Ga_{11}\equiv \Ga^0\Ga^1\cdots\Ga^9,
\eeqa
and the charge conjugation matrix $C$ as
\beqa
\ops_{\alpha}\equiv -\psi_{\beta}(C)_{\beta\alpha},\no
C^T=-C,~~C\Gamma^{\mu}C^{-1}=-(\Gamma^{\mu})^T,\no
C\Ga_{11}C^{-1}=-(\Ga_{11})^T.
\eeqa
Then the symmetry of matrices is the following:
\beqa
\mbox{Symmetric matrices}~&\cdots&~C\Ga^{\mu},C\Ga^{\mu\nu},
C\Ga^{\mu_1\cdots\mu_5},C\Ga_{11},C\Ga_{11}\Ga^{\mu},
C\Ga_{11}\Ga^{\mu_1\cdots\mu_4},\no
\mbox{Antisymmetric matrices}~&\cdots&~C,C\Ga^{\mu_1\mu_2\mu_3},
C\Ga^{\mu_1\cdots\mu_4},
C\Ga_{11}\Ga^{\mu\nu},C\Ga_{11}\Ga^{\mu_1\mu_2\mu_3}.
\eeqa
{}From this we can derive the following identities:
\beqa
\ops\Ga^{\mu}\ps=\ops\Ga^{\mu\nu}\ps=\ops\Ga^{\mu_1\mu_2\cdots\mu_5}\ps=0,\no
\ops\Ga^{11}\ps=\ops\Ga^{11}\Ga^{\mu}\ps
=\ops\Ga^{11}\Ga^{\mu_1\mu_2\cdots\mu_4}\ps=0.
\eeqa

\section{General Effective Action, 
Supersymmetry Transformations and 
Field Redefinitions}

Here we write all relevant terms which obey (\ref{condition2}) in the action, 
the supersymmetry transformations and the field redefinitions of 
$\ps(x),~T(x)$ and $A_{\mu}(x)$. The coefficients $f_i,~g_i,~h_i,
~j_i,~k_i$ and $l_i$ which appear in the latter equations are the nonsingular 
functions of $T$ around $T=0$, while $m_1$ is a constant.

The action:
\beqa
\label{action3}
S&=&-\sqrt{2}T_9\int d^{10}xe^{-\fr{1}{4}T^2}\no
&\times& [1+f_1(\ops\ps)+f_2(\ops\Ga^{\mu}\pa_{\mu}\ps)
+f_3(\ops\Ga_{11}\Ga^{\mu}\pa_{\mu}\ps)
+f_4(\pa^{\mu}T\pa_{\mu}T)+f_5F^{\mu\nu}F_{\mu\nu}\no
&&+f_{6}F^{\mu\nu}(\ops\Ga_{11}\Ga_{\mu\nu}\ps)
+f_{7}(\overline{\pa^{\mu}\ps}\pa_{\mu}\ps)
+f_{8}\pa^{\mu}T(\ops\pa_{\mu}\ps)
+f_{9}(\pa^{\mu}T\pa_{\mu}T)(\ops\ps)\no
&&+f_{10}\pa^{\mu}T(\ops\Ga_{11}\pa_{\mu}\ps)
+f_{11}F^{\mu\nu}(\ops\Ga_{\mu}\pa_{\nu}\ps)
+f_{12}F^{\mu\nu}(\ops\Ga_{11}\Ga_{\mu}\pa_{\nu}\ps)\no
&&+f_{13}F^{\mu\nu}(\ops\Ga_{\mu\nu\rho}\pa^{\rho}\ps)
+f_{14}F^{\mu\nu}(\ops\Ga_{11}\Ga_{\mu\nu\rho}\pa^{\rho}\ps)].
\eeqa

The supersymmetry transformations:
\beqa
\label{susyps3}
\de\ps_{\al}&=&g_1\ep_{\al}+g_2(\Ga_{11}\ep)_{\al}
+g_3\pa^{\mu}T(\Ga_{\mu}\ep)_{\al}+g_4\pa^{\mu}T
(\Ga_{11}\Ga_{\mu}\ep)_{\al}\no
&&+g_5F^{\mu\nu}(\Ga_{\mu\nu}\ep)_{\al}
+g_6F^{\mu\nu}(\Ga_{11}\Ga_{\mu\nu}\ep)_{\al}
+g_7(\pa^{\mu}T\pa_{\mu}T)\ep_{\al}\no
&&+g_8(\pa^{\mu}T\pa_{\mu}T)(\Ga_{11}\ep)_{\al}
+g_9\pa^2 T\ep_{\al}+g_{10}\pa^2 T (\Ga_{11}\ep)_{\al}\no
&&+g_{11}F^{\mu\nu}\pa_{\mu}T(\Ga_{\nu}\ep)_{\al}
+g_{12}F^{\mu\nu}\pa_{\mu}T(\Ga_{11}\Ga_{\nu}\ep)
+g_{13}\pa_{\mu}F^{\mu\nu}(\Ga_{\nu}\ep)_{\al}\no
&&+g_{14}\pa_{\mu}F^{\mu\nu}(\Ga_{11}\Ga_{\nu}\ep)_{\al}
+g_{15}F^{\mu\nu}\pa^{\rho}T(\Ga_{\mu\nu\rho}\ep)
+g_{16}F^{\mu\nu}\pa^{\rho}T(\Ga_{11}\Ga_{\mu\nu\rho}\ep)_{\al}, 
\eeqa
\beqa
\label{susyt3}
\!\!\! \de T&=&h_1(\oep\ps)+h_2(\oep\Ga_{11}\ps)
+h_3\pa^{\mu}T(\oep\Ga_{\mu}\ps)+h_4\pa^{\mu}T
(\oep\Ga_{11}\Ga_{\mu}\ps)\no
&&+h_5(\oep\Ga^{\mu}\pa_{\mu}\ps)
+h_6(\oep\Ga_{11}\Ga^{\mu}\pa_{\mu}\ps)
+h_7F^{\mu\nu}(\oep\Ga_{\mu\nu}\ps)\no
&&+h_8F^{\mu\nu}(\oep\Ga_{11}\Ga_{\mu\nu}\ps)
+h_9\pa^{\mu}T(\oep\pa_{\mu}\ps)
+h_{10}(\oep\pa^2\ps)\no
&&+h_{11}\pa^2 T(\oep\ps)+h_{12}\pa^{\mu}T\pa_{\mu}T(\oep\ps)
+h_{13}\pa^{\mu}T(\oep\Ga_{11}\pa_{\mu}\ps)
+h_{14}(\oep\Ga_{11}\pa^2\ps)\no
&&+h_{15}\pa^2 T(\oep\Ga_{11}\ps)
+h_{16}\pa^{\mu}T\pa_{\mu}T(\oep\Ga_{11}\ps)
+h_{17}F^{\mu\nu}\pa_{\mu}T(\oep\Ga_{\nu}\ps)\no
&&+h_{18}F^{\mu\nu}\pa_{\mu}T(\oep\Ga_{11}\Ga_{\nu}\ps)
+h_{19}\pa_{\mu}F^{\mu\nu}(\oep\Ga_{\nu}\ps)
+h_{20}\pa_{\mu}F^{\mu\nu}(\oep\Ga_{11}\Ga_{\nu}\ps)\no
&&+h_{21}F^{\mu\nu}(\oep\Ga_{\mu}\pa_{\nu}\ps)
+h_{22}F^{\mu\nu}(\oep\Ga_{11}\Ga_{\mu}\pa_{\nu}\ps)
+h_{23}F^{\mu\nu}\pa^{\rho}T(\oep\Ga_{\mu\nu\rho}\ps)\no
&&+h_{24}F^{\mu\nu}\pa^{\rho}T(\oep\Ga_{11}\Ga_{\mu\nu\rho}\ps)
+h_{25}F^{\mu\nu}(\oep\Ga_{\mu\nu\rho}\pa^{\rho}\ps)
+h_{26}F^{\mu\nu}(\oep\Ga_{11}\Ga_{\mu\nu\rho}\pa^{\rho}\ps),
\eeqa
\beqa
\label{susyA3}
\de A_{\mu}&=&j_1(\oep\Ga_{\mu}\ps)+j_2(\oep\Ga_{11}\Ga_{\mu}\ps).
\hspace{8cm}
\eeqa

Field Redefinitions:
\beqa
\label{fieldps3}
\ps^{\pr}&=&k_1\ps_{\al}+k_2(\Ga_{11}\ps)_{\al}
+k_3\pa^{\mu}T(\Ga_{\mu}\ps)_{\al}
+k_4\pa^{\mu}T(\Ga_{11}\Ga_{\mu}\ps)_{\al}\no
&&+k_5(\Ga^{\mu}\pa_{\mu}\ps)_{\al}
+k_6(\Ga_{11}\Ga^{\mu}\pa_{\mu}\ps)_{\al}
+k_7F^{\mu\nu}(\Ga_{\mu\nu}\ps)_{\al}
+k_8F^{\mu\nu}(\Ga_{11}\Ga_{\mu\nu}\ps)_{\al}\no
&&+k_{9}(\pa^2 \ps)_{\al}
+k_{10}(\Ga_{11}\pa^2\ps)_{\al}
+k_{11}\pa^{\mu}T (\pa_{\mu}\ps)_{\al}
+k_{12}\pa^{\mu}T (\Ga_{11}\pa_{\mu}\ps)_{\al}\no
&&+k_{13}\pa^2 T \ps_{\al}
+k_{14}\pa^2T (\Ga_{11}\ps)_{\al}
+k_{15}\pa^{\mu}T \pa_{\mu}T \ps_{\al}\no
&&+k_{16}\pa^{\mu}T \pa_{\mu}T (\Ga_{11}\ps)_{\al}
+k_{17}F^{\mu\nu}\pa_{\mu}T(\Ga_{\nu}\ps)_{\al}
+k_{18}F^{\mu\nu}\pa_{\mu}T(\Ga_{11}\Ga_{\nu}\ps)_{\al}\no
&&+k_{19}\pa_{\mu}F^{\mu\nu}(\Ga_{\nu}\ps)_{\al}
+k_{20}\pa_{\mu}F^{\mu\nu}(\Ga_{11}\Ga_{\nu}\ps)_{\al}
+k_{21}F^{\mu\nu}(\Ga_{\mu}\pa_{\nu}\ps)_{\al}\no
&&+k_{22}F^{\mu\nu}(\Ga_{11}\Ga_{\mu}\pa_{\nu}\ps)_{\al}
+k_{23}F^{\mu\nu}\pa^{\rho}T(\Ga_{\mu\nu\rho}\ps)_{\al}
+k_{24}F^{\mu\nu}\pa^{\rho}T(\Ga_{11}\Ga_{\mu\nu\rho}\ps)_{\al}\no
&&+k_{25}F^{\mu\nu}(\Ga_{\mu\nu\rho}\pa^{\rho}\ps)_{\al}
+k_{26}F^{\mu\nu}(\Ga_{11}\Ga_{\mu\nu\rho}\pa^{\rho}\ps)_{\al},
\eeqa
\beqa
\label{fieldt3}
T^{\pr}&=&T+l_1(\ops\ps)+l_{2}(\ops\Ga^{\mu}\pa_{\mu}\ps)
+l_{3}(\ops\Ga_{11}\Ga^{\mu}\pa_{\mu}\ps)+l_4(\pa^{\mu}T\pa_{\mu}T)
+l_5\pa^2 T\no
&&+l_6F^{\mu\nu}F_{\mu\nu}
+l_{7}F^{\mu\nu}(\ops\Ga_{11}\Ga_{\mu\nu}\ps)
+l_{8}(\ops\pa^2\ps)
+l_{9}(\overline{\pa^{\mu}\ps}\pa_{\mu}\ps)\no
&&+l_{10}\pa^{\mu}T(\ops\pa_{\mu}\ps)
+\l_{11}\pa^2T(\ops\ps)+l_{12}\pa^{\mu}T\pa_{\mu}T(\ops\ps)\no
&&+l_{13}(\ops\Ga_{11}\pa^2\ps)
+l_{14}\pa^{\mu}T(\ops\Ga_{11}\pa_{\mu}\ps)
+l_{15}F^{\mu\nu}(\ops\Ga_{\mu}\pa_{\nu}\ps)\no
&&+l_{16}F^{\mu\nu}(\ops\Ga_{11}\Ga_{\mu}\pa_{\nu}\ps)
+l_{17}F^{\mu\nu}(\ops\Ga_{\mu\nu\rho}\pa^{\rho}\ps)
+l_{18}F^{\mu\nu}(\ops\Ga_{11}\Ga_{\mu\nu\rho}\pa^{\rho}\ps)\no
&&+l_{19}F^{\mu\nu}\pa^{\rho}T(\ops\Ga_{\mu\nu\rho}\ps)
+l_{20}F^{\mu\nu}\pa^{\rho}T(\ops\Ga_{11}\Ga_{\mu\nu\rho}\ps).
\eeqa
\beqa
A_{\mu}' &=& m_1 A_{\mu}.
\hspace{11.0cm}
\eeqa

\end{document}